# A New Peak Detection Method for Single or Three-Phase Unbalanced Sinusoidal Signals

Jusong Rim[1,3], Cholyong Ri[1], Hongchol Jin[2], Choljin Ohr[3], Choljun Rim[3], Hyewon Ri[2]

1. Department of Physics, KimIlSung University, Pyongyang, DPR of Korea
2. Department of Thermal Engineering, KimChaek University of technology, Pyongyang, DPR of Korea
3. Department of Control Science, University of Sciences, Pyongyang, DPR of Korea

*Abstract-* **In this paper, a fast amplitude detection method for the single or three-phase unbalanced sinusoidal is reported. The proposed method is a method of the amplitude detection for a single phase or three phase unbalanced sinusoidal signal, based on detecting of the pulse width corresponding to the peak amplitude. The detection period is the half period of the input signal. This method is independent of the three-phase signal's unbalance and phase sequence. The proposed method was verified through experiments for the single or three-phase unbalanced signals.**

*Index Terms*—**Amplitude detection, detection, fast amplitude detection, unbalanced three-phase signals**

## I. INTRODUCTION

THE developing of electrical/electronic technology, it's necessary to estimate more quickly electrical parameters in single or three-phase applications. The fundamental voltage, current, and phase angle are very important in power system. The sinusoidal amplitude detector is broadly used in electrical processes such as power system monitoring, protection and fault detection. In particular, voltage sag detecting techniques include the root mean square (RMS), Fourier transform, and peak voltage detection methods [10].

In some cases, three-phase unbalance is occurred because of several reasons in the three-phase system, so in real application process needs fast peak detection for the three-phase unbalanced signal.

Several methods [1, 2] proposed to improve the response time of single-phase sinusoidal amplitude detection. Some researches [8, 9, 16] proposed the peak amplitude detection method by the using of phase-locked loop. And also there are rapid peak detection methods based on a nonlinear adaptive filter [10], a modified Prony's model [15], and fast voltage sag detection method based on peak detection [19].

Three-phase systems are widely used in the industry power system. The three-phase peak detectors have been developed [3,4], which can measure the peak value of three-phase sinusoidal signals by the using of three phases of voltage. The high speed measure method by the using of two phases of voltage was proposed [5, 6]. However, these methods can only be used in balance state and has deviation in unbalance state.

Recently, several techniques have been proposed the fast detection methods of amplitude, phase and unbalance parameters in three-phase systems, such as three-phase synthetic testing with asymmetrical short-circuit currents[7], three-phase three-angle algorithm (TP–TA)[11], method using a modified all-pass filter (MAPF) [12], method for the line-interactive dynamic voltage restorer (DVR)[13], phase-locked-loop (PLL) circuit based on the p-q theory[14], low-pass notch filter PLL (LPN-PLL)[17], Maximum Likelihood Estimators (MLEs)[18].

In this paper, a new peak detection method for the single or three-phase unbalanced sinusoidal signal is proposed. The proposed method is based on detecting of the pulse width corresponding to the peak amplitude. The detection period is the half period of the input signal. And the proposed method is independent of the phase sequence and unbalanced condition and can be also used in the single or three -phase three wire and three-phase four wire. The performance of this method is demonstrated by a real electronic circuit.

## II. METHOD OF DETECTING THE PEAK AMPLITUDE

The sinusoidal signal is generally signed as follows.

$$V(t) = V_p \sin(\omega t) \qquad (1)$$

, where $V(t)$ is instantaneous value of input signal，$V_p$ is the maximum amplitude and $\omega$ is the angular frequency of the sinusoidal signals. From here, $V_p$ is

$$V_p = \frac{V(t)}{\sin(\omega t)} \qquad (2)$$

If we fix $V(t)$ at the measure moment as $V(t) = V_h$ in the measure circuit, $V_p$ will be indicated as follows

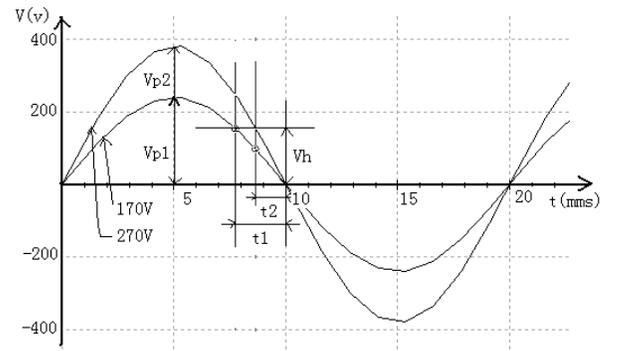

Fig. 1. In the case of the same frequency and the phase change of the same voltage line variation.

$$V_p = \frac{V_h}{\sin(\omega t_h)} \qquad (3)$$

In where the $V_h$ will be a constant dependent on the measure circuit, $t_h$ is the time that input signal takes from zero cross point to $V_h$ value. From ω=2πf, if we know the frequency $f$ of system and time value $t_h$, we can easily calculate $V_p$.

Fig. 1 shows the change of $t_h$ in the sinusoidal signal of the different amplitude in the fixed frequency.

Using comparator can realize the amplitude fixing of input signal.(Fig. 2)

The signal obtained from multiplier process of the up half period and down one of input signal is (b) in Fig. 4. This pulse width flows in proportion to $t_h$.

In case of three-phase signal, pulses of every phase signal pass the add circuit. The output signal is the pulses row as the output of Fig. 10 because the zero-cross moments of every phase signal are different.

From the above, we don't use the rectifier and AD converter.

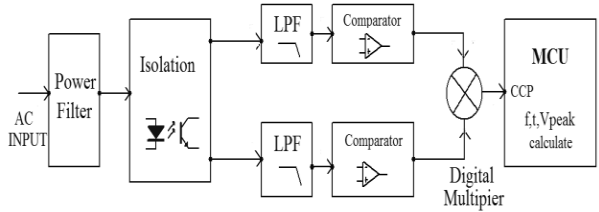

Fig. 2. Block diagram of the proposed method in the single phase.

And we detect $f$ and $t_h$ of every half period to calculate the peak amplitude.

Due to the independence of measure, peak amplitude value of three phase signal is not influenced from the other phase in the measure and this measure method can be also used in the three phase three wire and three phase four wire.

## III.  HARDWARE IMPLEMENTATION

To implement the proposed method written as equation (3) by an electronic circuit, we have to measure $f$ and $t_h$ together at a half period of sinusoidal signals. $t_h$ appears twice at each half period signal (the begin and the end of half period signal) and with $t_h$ of the before half period signals and later one. And have to detect the zero cross. Thus we compose the input circuit with an opt coupler to isolate input/output, improve the repeat characteristic without detecting of the zero cross.  The input resistor and the collector's bias resistor of the optocoupler are selected by the input voltage range and the linear character of the optocoupler. Finally, a pulse with $t_h$ at $V_h$ is obtained as result of multiplying two signals from the analog comparator connected with the optocoupler.

The threshold of the comparator is in the linear range of the optocoupler's output signal.  We input this signal to the capture input of the micro controller and measure $f$ and $t_h$, calculate $V_p$ value. (Fig.3, 4)

Fig.5 shows the detection block of the peak amplitude in the three phase sinusoidal signal .This method can apply to all of three phase three wire and three phase four wire with change of the input isolation circuit.

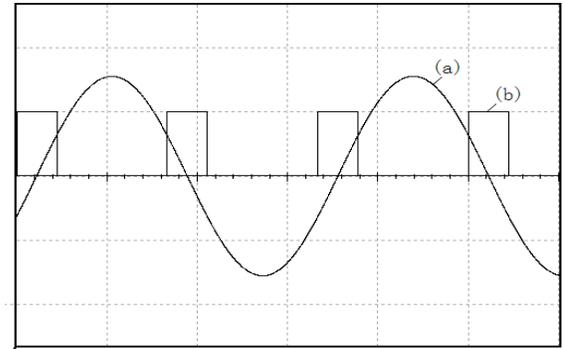

ch a: 200/div, ch b: 5V/div ,5mV/div

Fig. 4. The relation between input sinusoidal signal and output pulse signal.

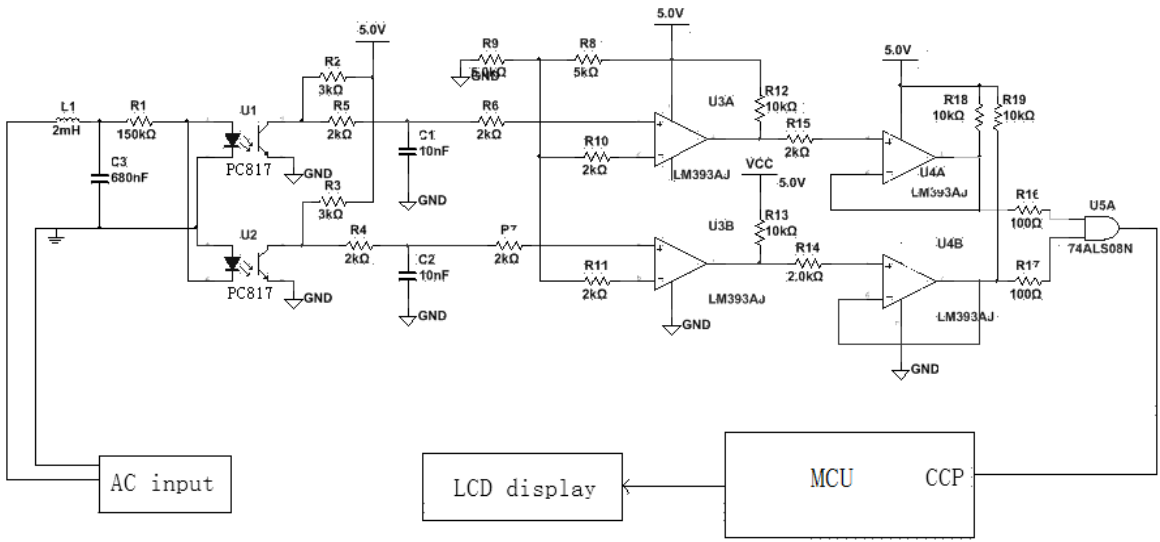

Fig. 3. Detect block of the single phase sinusoidal amplitude

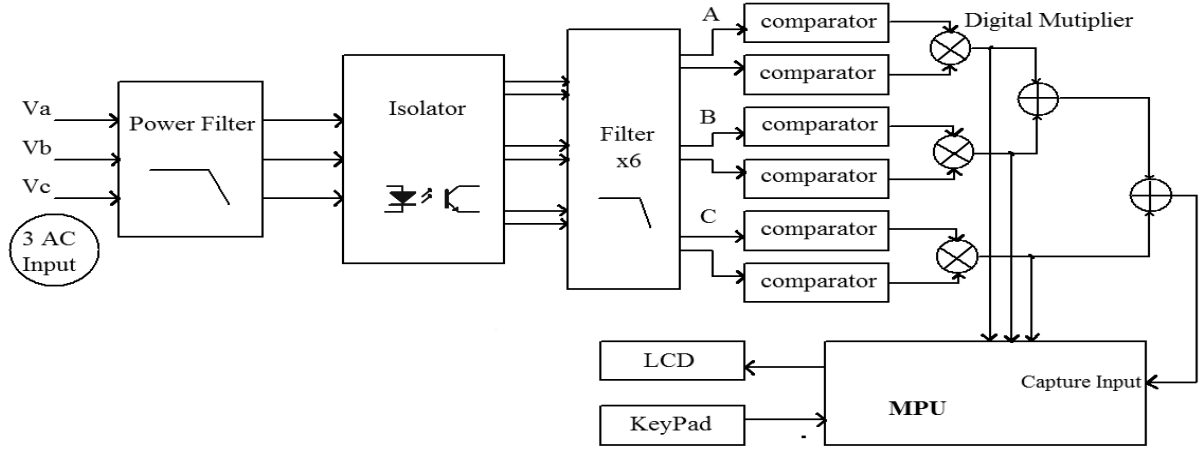

Fig. 5. Detect block of the three phase sinusoidal amplitude

The part of main circuit is same as a single phase detecting circuit.

But the output pulse of obtained each phase, after is processed by the adder circuit, is inputted to the capture input of processor.

Also input pulses of three phase signal input to the processor to identify the phase, which make it possible to calculate voltage and phase angle of each phase respectively.

If a detecting circuit is composed as this, we can calculate the amplitude of each phase signal and the unbalance condition of input signal，identify the phase sequence change, the phase cutting, the phase short from detecting of the phase angle variation.

## IV. TEST RESULT

### A. Linearity

To test the linearity of the proposed detector, we tested a single-phase voltage. Fig.6 shows the test result. The solid line is theory value, and the dotted line is the measured result.。

To decrease the calculating time, use the approximation. From $\omega t \ll 1$, (3) is changed as

$$V_n = \frac{V_h}{\sin(\omega t_h)} \approx \frac{V_h}{\omega t_h} \qquad (4)$$

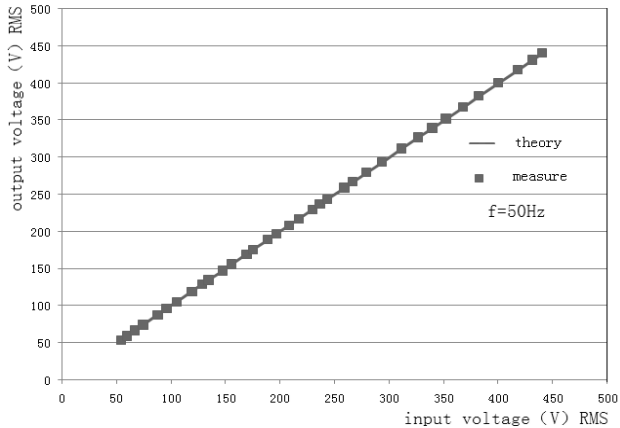

Fig. 6. The linearity of the proposed method

$$V_p = V_n + \frac{A}{V_n} \qquad (5)$$

, where $A$ is a constant.

Although the use of approximate calculations, the detector appears good results in the input voltage of 50 ~ 440V range. The error voltage is less than 0.2V in the most of range and is less than 0.5V in the boundary point. The error mostly relates with the accuracy of frequency detecting in the processor and the approximation.

This shows that the proposed detector has the sufficient linearity in real application.

### B. Frequency Characteristic

Fig. 7 shows the test result of frequency characteristic in the proposed detector. The frequency of input signal is 31~300Hz and is tested in the certain voltage. Test result shows that there is not the change of voltage in the error range. The reason of error is same as before.

This shows that the proposed method is independent of the frequency characteristic.

### C. Effect of phase sequence

Fig. 8 shows the wave form of output signal in the proposed method when the phase sequence is changed. The phase sequence of Fig. 8(a) is a→b→c and Fig. 8(b) is b→a→c. As known as Fig. 8, in the case of the change of the phase sequence, the sequence of pulse row inputting to capture input of the processor is changed, but there is not a variation of the peak

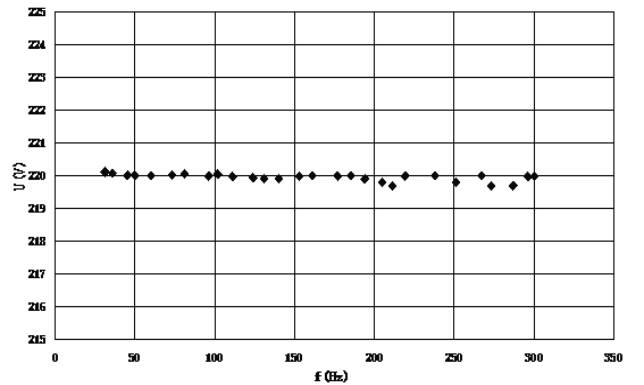

Fig. 7. The frequency characteristic of the proposed method

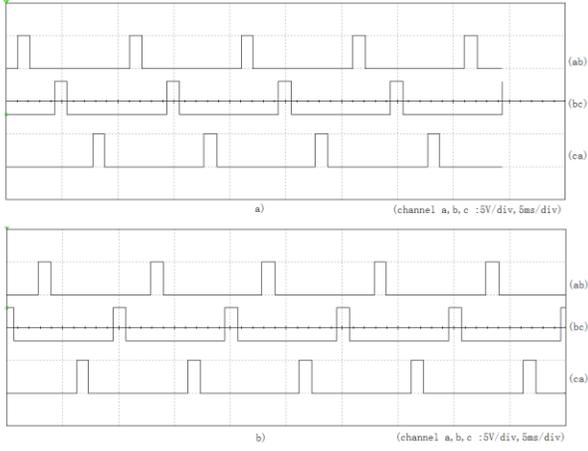

Fig. 8. The effect of the phase sequence in three phase signal detecting.

amplitude in the corresponding phase.

Because the proposed method determinates the peak amplitude value from checking of the sequence of inputting pulse arrange, immediately detects the change of the phase sequence.

The result shows that the proposed method is independent of the phase sequence and can detect the change of phase sequence together.

### D. Response Characteristic

The proposed method can detect the corresponding peak amplitude of phase in each output pulse. Accordingly the detection period is the half period of the input signal. (Fig. 4)

### E. Effect of Harmonics

Fig. 9 shows the test result of the effect of harmonics. (a) is the output pulse signal without harmonics, (b) is with harmonics, (c) is with using of LPF in the input of detector and analog comparator. In this method, the removal of harmonics' effect is important because we calculate the peak amplitude from measuring the output pulse width.

This problem can be solved by using of a pre-filter in the input of this detector and a low-pass filter in the input of analog comparator, the software process.

This paper proposes the technique that can cancel the effect of waveform distortion about the zero cross by measuring the interval between the times when the signal reaches the threshold in the first and the second half-periods, but not by measuring the interval between the zero cross and the time when the signal reaches the threshold in each half-period. (Fig.4) In the proposed method a pre-filter and an active low-pass filter are located at the before and after of the optocoupler to cancel the effect of the harmonics, uses the digital processing to cancel the parasitic vibration occurred at the before and after of the main pulse waveform. If the harmonic distortion is too large, an active power filter may be inserted into the input to cancel the harmonics and distortions.

### F. Effect of unbalance

When the three-phase system is unbalance, it may include phase unbalance, amplitude unbalance, or both.

In amplitude unbalance, three-phase signal can be represented as

$$V_A(t) = (V_P + \Delta V_1)\sin(\omega t) \qquad (6)$$
$$V_B(t) = (V_P + \Delta V_2)\sin(\omega t - 120°) \qquad (7)$$
$$V_C(t) = (V_P + \Delta V_3)\sin(\omega t + 120°) \qquad (8)$$

, where $\Delta V_1$, $\Delta V_2$ and $\Delta V_3$ are three-phase amplitude error signal. In this method, $t_h$ value is according to amplitude value in each phase signal.(Fig. 10) Thus, this method can detect the unbalanced amplitude of each phase respectively and detect the unbalanced state.

In phase unbalance,

$$V_A(t) = V_P \sin(\omega t + \phi_1) \qquad (9)$$
$$V_B(t) = V_P \sin(\omega t - 120° + \phi_2) \qquad (10)$$
$$V_C(t) = V_P \sin(\omega t + 120° + \phi_3) \qquad (11)$$

, where $\phi_1$, $\phi_2$ and $\phi_3$ are each phase deviations.

In three-phase 3 wire, $V_A(t), V_B(t), V_C(t)$ are signals between AB, BC, CA, the phase unbalance is represented as amplitude change of (6), (7), (8).

In three-phase 4 wire, because the deviation value is included to $t_h$ value, amplitude change is not matter to phase unbalance.

This shows that the proposed method is independent of the three-phase unbalance.

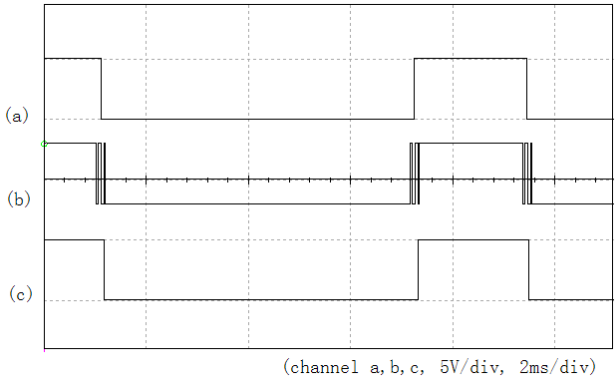

Fig. 9. Effect of harmonics in the proposed method

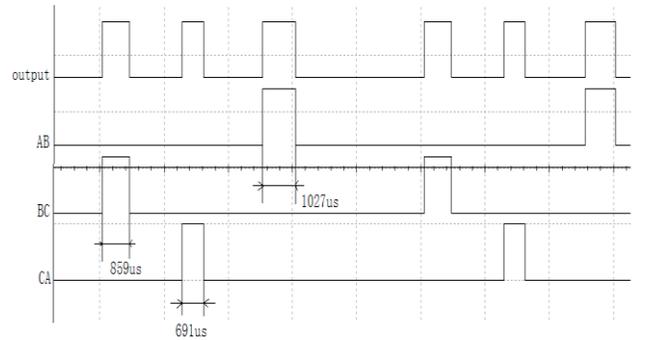

Fig. 10. Variant of output pulse width in unbalance condition

## V. Conclusion

In this paper, a fast amplitude detection method for the single or three-phase unbalanced sinusoidal signal was presented.

The proposed method is a method of the amplitude detection for a single phase or three phase unbalanced sinusoidal signal, based on detecting of the pulse width corresponding to the peak amplitude. The detection period is the half period of the input signal.

In this method, the single phase peak detection method is applied to the peak detection of three-phase sinusoidal signal.

This method is independent of the phase sequence and can detect the change of phase sequence immediately. And this method is independent of the three-phase signal's unbalance.

The proposed method has to detect the frequency of the input signal and needs analog filter or digital filter because it depends on the signal harmonics.

## Acknowledgment

The author wishes to thank the IEEE for providing this template and all colleagues who previously provided technical support.